\documentclass[journal]{IEEEtran}

\usepackage[colorlinks,urlcolor=blue,linkcolor=blue,citecolor=blue]{hyperref}

\usepackage{color,array}

\usepackage{graphicx}

\usepackage{amsmath,amsfonts,bbold}
\usepackage[utf8]{inputenc}
\usepackage[T1]{fontenc}
\usepackage{algorithmic}
\usepackage{algorithm}
\usepackage{textcomp}
\usepackage{stfloats}
\usepackage{url}
\usepackage{multirow,booktabs}
\usepackage{verbatim}
\usepackage{graphicx}
\usepackage{cite}
\usepackage{cleveref}
\usepackage[normalem]{ulem} 
\usepackage{xcolor}

\usepackage[caption=false,font=normalsize,labelfont=sf,textfont=sf]{subfig}

\begin{document}

\title{Hardware-Algorithm Co-Optimization of Early-Exit Neural Networks for Multi-Core Edge Accelerators} 

\author{Alaa Zniber,~\IEEEmembership{Member,~IEEE,} Arne Symons,~\IEEEmembership{Member,~IEEE,} Ouassim Karrakchou,~\IEEEmembership{Senior Member,~IEEE} \\Marian Verhelst,~\IEEEmembership{Fellow,~IEEE,} Mounir Ghogho,~\IEEEmembership{Fellow,~IEEE}
\thanks{This research has received funding from the European Union’s Horizon
research and innovation program under grant agreement No 101070374. \\ \indent Zniber A., Karrakchou O., Ghogho M. are with the TICLab, International University of Rabat, Morocco (e-mail:
alaa.zniber@uir.ac.ma; ouassim.karrakchou@uir.ac.ma; mounir.ghogho@uir.ac.ma) \\ \indent Symons A., Verhelst M. are with MICAS, KU Leuven, Belgium (e-mail: arne.symons@kuleuven.be,  marian.verhelst@kuleuven.be)}}



\maketitle

\begin{abstract}
The deployment of Early-Exiting Neural Networks (EENNs) on edge accelerators requires optimizing not only the network architecture but also its hardware deployment. Exit configuration, quantization, and hardware workload mapping interact in non-trivial ways, influencing memory traffic, accelerator utilization, and ultimately the energy-latency trade-off. This work presents a hardware-aware co-design framework for EENNs that jointly optimizes exit configuration, quantization-aware training, and multi-core hardware mapping within a unified NAS process. Leveraging analytical design space exploration, the framework identifies efficient workload mappings for each candidate architecture while providing accurate latency and energy estimates during the search. We further formulate EENN deployment as a constrained multi-objective optimization problem balancing predictive accuracy, energy-latency product, exit overhead, and dynamic inference efficiency. Experimental results on CIFAR-10 demonstrate that the proposed framework achieves over a 50\% reduction in energy-latency product compared with static baselines under 8-bit quantization. These results demonstrate that jointly optimizing architecture and deployment is essential for realizing the full efficiency potential of dynamic inference on heterogeneous edge accelerators.
\end{abstract}



\maketitle

\section{Introduction}
\label{intro}

The rapid advancement of high-performance processing devices and cloud technologies has driven a notable increase in the architectural complexity of Deep Learning (DL) models, leading to significant performance gains in various domains, such as large language models \cite{llm_compute}. However, deploying DL models at the edge is often necessary to comply with data privacy regulations (e.g., in healthcare applications) \cite{Li2024} or to meet real-time processing requirements (e.g., in autonomous driving) \cite{low_latency_survey}. Additionally, the substantial energy consumption of large-scale DL models in the cloud raises serious environmental concerns \cite{green_ai}. Consequently, there is an urgent need to reduce the computational complexity of DL models, making them more suitable for resource-constrained edge hardware while enhancing their energy efficiency \cite{algo_hardware}.

\begin{figure}[t]
    \centering
    \includegraphics[width=0.9\columnwidth]{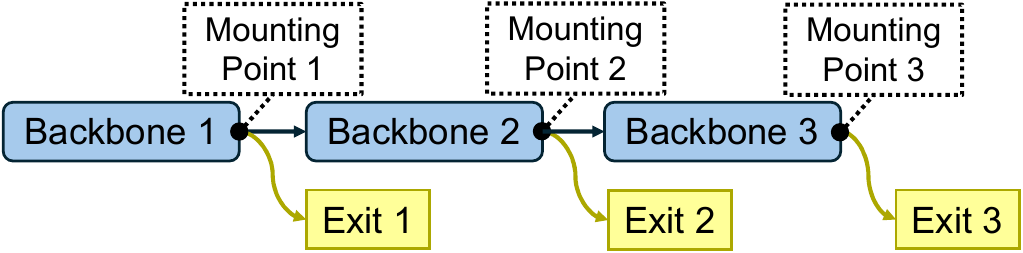}
    \caption{An example early exiting network with 3 backbone blocks and 3 exits.}
    \label{fig:eenn}
\end{figure}

In this context, Dynamic Neural Networks (DyNNs) have emerged as a promising solution to this challenge by adapting their computational effort to the complexity of each input sample rather than executing a fixed inference pipeline \cite{squeeze, deformable, dynamic_relu}. Unlike conventional static networks, DyNNs dynamically adjust their execution, allocating additional computation only when necessary. This adaptive inference paradigm enables substantial reductions in latency and energy consumption while maintaining predictive performance \cite{dynamic_routing, skipping_layers}. Among these approaches, Early Exiting Neural Networks (EENNs) have attracted significant attention due to their simplicity and effectiveness \cite{survey_early_exit}. In the case of classification, an EENN consists of a backbone network augmented with intermediate classifiers (ICs) positioned at predetermined exit points (cf. \Cref{fig:eenn}). During inference, an input sample is processed sequentially through the network. At each exit point, an IC evaluates the sample before further processing in the backbone. The classification output (i.e., the highest class probability) is then compared against a user-defined confidence threshold. If the probability exceeds the threshold, the EENN confidently returns the classification result and halts further computation. Otherwise, feature extraction continues in the backbone until the next exit point is reached.

Although EENN can effectively reduce computational complexity, the resulting energy savings are strongly dependent on the characteristics of the target hardware platform. While early-exit networks have been studied primarily from an algorithmic perspective, their deployment on heterogeneous edge accelerators introduces additional structural constraints. Exit placement changes activation tensor dimensions and intermediate memory footprints, quantization affects representational capacity and early-exit confidence behavior, and accelerator mapping influences inter-core communication and memory reuse. Despite these hardware-dependent effects, most existing Neural Architecture Search (NAS) approaches optimize EENNs using proxy metrics such as MAC operations or architecture-only search formulations \cite{edanas, nachos}. Although these metrics correlate with computational complexity, they often fail to reflect the true cost of edge deployment, where memory transfers and data movement frequently dominate energy consumption. To address this limitation, some recent studies, such as \cite{hadas}, incorporate hardware-in-the-loop evaluation to estimate latency and energy consumption using physical measurements. While this approach improves the accuracy of hardware cost estimation, it lacks scalability and flexibility because the search process remains tightly coupled to a specific physical platform. Furthermore, hardware-in-the-loop evaluation limits systematic exploration of alternative accelerator configurations and mapping strategies. These limitations indicate that optimizing EENNs requires a deployment-aware design methodology that explicitly models the interaction between network architecture, dynamic inference behavior, and hardware resource allocation, rather than relying solely on proxy metrics or treating hardware evaluation as a post hoc validation step.

Additionally, modern edge platforms are increasingly heterogeneous, integrating multiple specialized processing elements and hardware accelerators for different operations (e.g., convolutions \cite{conv_accelerate} and softmax \cite{softmax_accelerate}). Consequently, efficiently mapping EENN components across heterogeneous multi-core architectures is essential to avoid load imbalance, excessive inter-core communication, and unnecessary memory transfers that can offset the benefits of early exiting. To the best of our knowledge, the interaction between EENN execution behavior and heterogeneous multi-core resource allocation has not been systematically explored.

To address this challenge, we introduce a hardware-aware optimization framework for EENN that explicitly integrates quantization effects and hardware resource allocation into both architecture design and training. Rather than relying on proxy metrics or post hoc hardware evaluation, our approach embeds hardware considerations directly within the search process to ensure compliance with modern edge accelerator constraints. The framework extends conventional NAS pipelines with components tailored to dynamic inference on heterogeneous edge platforms. First, recognizing the ubiquity of low-precision arithmetic in edge deployment, we adopt quantization-aware training to account for precision-induced shifts in exit behavior and predictive performance. Second, hardware mapping and resource allocation are performed using the Stream design space exploration framework \cite{stream}, which analytically explores mappings across heterogeneous multi-core accelerators and configurable hardware components. This joint exploration provides accurate energy and latency estimates while preserving flexibility across diverse hardware platforms. Third, we refine the search space by introducing exit-dependent structural constraints that explicitly limit intermediate overhead and promote effective early exiting, encouraging samples to terminate in earlier layers whenever appropriate.

The contributions of this work are as follows:
\begin{itemize}
\item We provide a systematic characterization of the interaction between quantization, exit placement, and multi-core accelerator mapping in EENN, demonstrating that minor architectural variations can produce significant hardware-level performance differences.
\item We formulate early-exit network deployment as a constrained multi-objective co-design problem that jointly accounts for accuracy, energy-latency product, exit overhead, hardware mapping, and dynamic inference efficiency.
\item We develop a hardware-aware optimization framework integrating quantization-aware training with analytical design space exploration to navigate this deployment-aware search space.
\item We validate the approach on a quad-core edge TPU model, demonstrating substantial improvements in energy-latency efficiency relative to static baselines.
\end{itemize}

The remainder of the paper is structured as follows. \Cref{related_work} surveys relevant literature. In \Cref{study}, we examine the complex interactions between quantization, EENN mounting points, and accelerator types to motivate our hardware-algorithm co-optimization framework. Then, our proposed hardware-aware NAS for EENN is presented in \Cref{approach}. Experimental evaluation and discussion are provided in \Cref{exprms}. Finally, \Cref{conclusion} concludes the paper.

\section{Related Work}
\label{related_work}

Existing work on hardware-efficient deep learning can be broadly divided into two complementary research directions. Design Space Exploration (DSE) focuses on optimizing the deployment of neural network workloads on hardware accelerators, whereas hardware-aware Neural Architecture Search (NAS) optimizes network architectures while considering hardware objectives. We review both directions in the following subsections.

\subsection{Design Space Exploration}

Hardware accelerators are specialized computing architectures designed to execute DL workloads more efficiently than general-purpose processors by improving performance and energy efficiency, particularly in resource-constrained edge environments. To fully exploit these architectures, Design Space Exploration (DSE) frameworks have become an essential tool for evaluating and optimizing the interaction between neural network workloads and hardware platforms. In general, DSE addresses two complementary optimization problems: (i) hardware design exploration, which searches for efficient accelerator architectures (e.g., number of cores, processing-element array dimensions, or memory hierarchy); and (ii) mapping exploration, which determines efficient workload mappings onto a given hardware platform through scheduling, tiling, dataflow selection, and workload allocation across available processing resources. Throughout this process, DSE frameworks provide accurate estimates of hardware metrics, such as latency and energy consumption, to guide optimization.

Several DSE frameworks have been proposed to optimize DL deployment on hardware accelerators. Timeloop~\cite{timeloop} explores mappings of DL workloads onto accelerator architectures by optimizing dataflows and memory hierarchies. ZigZag~\cite{zigzag} further extends the mapping design space by supporting uneven mapping schemes and heterogeneous memory hierarchies through an analytical performance model. However, both frameworks are limited to single-core accelerators. Stream~\cite{stream} extends DSE to heterogeneous multi-core accelerator architectures by jointly exploring layer-to-core assignments, dataflows, and memory allocation while explicitly modeling off-chip and inter-core communication. In addition to identifying an efficient workload mapping, Stream provides analytical estimates of latency and energy consumption for each explored deployment.

In this work, the hardware architecture is assumed to be fixed, and Stream is integrated into the NAS framework to optimize the deployment of every candidate EENN. Specifically, Stream determines the hardware mapping and workload allocation that minimize the deployment cost of each candidate architecture while providing accurate latency and energy estimates to guide the search. This enables hardware mapping and multi-core workload allocation to be jointly optimized with the EENN architecture during NAS.

\subsection{Hardware-aware Neural Architecture Search}

As deep neural networks have increasingly been deployed on resource-constrained edge devices, NAS has evolved from optimizing predictive accuracy alone to jointly considering hardware objectives such as inference latency, energy consumption, memory footprint, and hardware area. Conventional hardware-aware NAS incorporates these objectives either as hard constraints that eliminate infeasible architectures \cite{ofa} or as objectives within joint or multi-objective optimization formulations \cite{joint_nas, nsganet}.

Hardware-aware NAS has also been extended to automate the design of EENNs. Existing approaches optimize exit placement while balancing predictive accuracy and computational efficiency through reinforcement learning \cite{s2dnas}, evolutionary algorithms \cite{enas4d}, and multi-objective search formulations \cite{edanas}. To estimate hardware efficiency during the search, these methods typically rely on architecture-level proxy metrics such as MAC counts \cite{edanas, nachos}, learned performance predictors \cite{enas4d}, or direct hardware measurements obtained through hardware-in-the-loop evaluation \cite{hadas}. Overall, existing hardware-aware NAS methods optimize architecture design and deployment largely in isolation. Consequently, they fail to capture the interaction between EENN architecture, dynamic inference, and hardware execution, limiting deployment flexibility and the achievable accuracy-energy-latency trade-offs.

In contrast, our framework extends the optimization beyond exit configuration by jointly considering quantization-aware training, hardware mapping, and multi-core workload allocation within the architecture search process. In addition, it incorporates EENN-specific deployment objectives, including exit overhead and dynamic inference efficiency, allowing candidate architectures to be evaluated under realistic deployment conditions throughout the search. By integrating architectural and deployment-level decisions into a unified optimization framework, the proposed approach identifies architectures that achieve a more favorable trade-off between predictive performance and hardware efficiency.

\section{Characterization of Model-Hardware Interaction in EENN}
\label{study}

In this section, we investigate the impact of hardware on EENN design across several dimensions. First, we explore the effect of quantization on the performance of a fixed EENN, considering accuracy, energy consumption, and latency. Second, we examine how different EENN architectures derived from the same backbone behave on specific hardware. Third, we evaluate the deployment of EENN architectures on different homogeneous and heterogeneous accelerators. Our goal is to show to what extent variations in EENN models interact with hardware architecture, leading to differences in performance.

\subsection{Methodology of EENN Performance Evaluation}
\label{hw_section}

We conduct our study on an image classification task using \textit{CIFAR-10} \cite{cifar}, a well-known dataset from the MLPerf Tiny benchmark \cite{mlperf}, which is dedicated to edge devices and applications. The evaluated metrics include model accuracy and hardware costs (i.e., energy and latency).
Formally, we define the exit ratio of the $i^{\text{th}}$ early exit, $\text{ER}_i$, as the proportion of input data that exit at $i$, satisfying a predefined confidence threshold. Thus, the average accuracy and energy-latency product of the full EENN can be expressed as follows:
\begin{align}
\label{eqn:acc}
&\text{ACC\_{avg}} = \sum_{i=1}^m \text{ER}_i  \times \text{ACC}_i \\
&\text{ET\_{avg}} = \sum_{i=1}^m \text{ER}_i \times \text{ET}_i
\end{align}

\noindent where $\text{ACC}_i$ is the accuracy of the samples that exited at point $i$ and $\text{ET}_i$ is the product of energy ($E$, in joule) and latency/delay ($T$, in cycles) of the $i^{\text{th}}$ subnetwork (i.e. from the input up to the considered exit). 

We evaluate our hardware cost using an enhanced version of the \textit{Stream} DSE framework \cite{stream}, which provides detailed energy and latency estimates while accounting for dynamicity and quantization. Stream considers factors such as off-chip memory usage and Network-on-Chip (NoC) core-to-core communication overhead for both activations and weights—critical for understanding system-level performance and energy efficiency. This evaluation includes the computational cost of each backbone layer and the overhead introduced by early exit blocks, offering insights into the trade-offs between energy consumption and latency. Moreover, Stream supports optimized workload mapping onto the multi-core architecture through an intra-core \textit{temporal mapping} optimization engine called LOMA \cite{loma} and an inter-core \textit{workload allocation} GA-based engine \cite{ga_inter_core}. Once the best workload allocation for a given multi-core accelerator is determined, we use Stream’s hardware cost breakdown to extract relevant hardware metrics—namely, the energy $E_k$ and latency $T_k$ for a layer $k$. The energy for a given layer $k$ is defined as follows:
\[
E_k = \text{Computation}_k + \text{NoC Traffic}_k
\]
where the computational cost is combined with the traffic across the NoC to estimate the overall utilization of the resources of the considered layer, while latency $T_k$ is defined as the duration of executing the layer $k$. Therefore, for a sub-network $i$ (e.g. all layers up to exit $i$), we can define $\text{ET}_i$ as follows:
\begin{equation}
\label{over}
    \text{ET}_i = \left( \sum_{k=1}^i E_k \right) \times \left( \sum_{k=1}^i T_k \right)
\end{equation}
It is worth mentioning that $\text{ET}_i$ includes the overhead of all previous intermediate exits, as all of them are executed during inference before a sample exits at exit $i$.

\subsection{Impact of Quantization}
To study the impact of quantization, we use a modified MobileNetV2 backbone with a quad-core Edge TPU target (cf. Appendix A for more details). We enhance the backbone with three intermediate exit points placed at approximately $12\%$, $25\%$, and $60\%$ of the total MAC operations. We fine-tune the confidence threshold $\tau$ between $80\%$ and $95\%$ for all exit points.
We reduce the precision from 32-bit floating point for weights and activations to 8-bit and 4-bit integer precision, as these are the most widely used quantization levels in edge accelerators. Furthermore, to assess the accuracy improvements contributed by each component (i.e., backbone or early exits), we apply a heterogeneous mixed-precision quantization scheme that disentangles the precision of the backbone and early exits.
To train our model, we adopt a quantization-aware training scheme \cite{survey_efficient} for greater flexibility in representation learning and minimal sensitivity to quantization noise (cf. Appendix B).

\begin{table*}[t!]
    \centering
    \caption{Accuracy and exit ratio for different quantization configurations. FP/INT $X+Y$ means $X$-bit precision for the backbone and $Y$-bit precision for the exit point classifiers. Cum. Params denotes cumulative number of parameters, Cum. MACs denotes cumulative multiply-accumulate operations.}
    \begin{tabular}{ccc|cc|cc|cc|cc}
        \toprule
        & \multicolumn{2}{c}{\textbf{Exit 1 (D)}} & \multicolumn{2}{c}{\textbf{Exit 2 (F)}} & \multicolumn{2}{c}{\textbf{Exit 3 (I)}} & \multicolumn{2}{c}{\textbf{Exit 4 (K)}} & &  \\
        \hline
        Cum. Params & \multicolumn{2}{c}{30.922} & \multicolumn{2}{c}{71.946} & \multicolumn{2}{c}{354.634} & \multicolumn{2}{c}{1.439.498} & &  \\
        \hline
       Cum. MACs & \multicolumn{2}{c}{24.515.584} & \multicolumn{2}{c}{48.752.640} & \multicolumn{2}{c}{118.307.840} & \multicolumn{2}{c}{195.377.152} & &  \\
         \hline  \hline
        \cmidrule(lr){2-3}\cmidrule(lr){4-5}\cmidrule(lr){6-7}\cmidrule(lr){8-9}
        & Accuracy & Exit Ratio & Accuracy & Exit Ratio  & Accuracy & Exit Ratio& Accuracy & Exit Ratio & ACC\_{avg} & ET\_{avg} \\
        \cmidrule(lr){2-3}\cmidrule(lr){4-5}\cmidrule(lr){6-7}\cmidrule(lr){8-9}
        \hline
        FP32+32 & 98.48 & 34.21 & 94.73 & 14.22 & 93.80 & 28.22 & 63.68 & 23.35 & 88.50 & 4921 \\
        \cline{1-11}
         INT8+8  & 99.10 & 25.49 & 96.86 & 15.31 & 95.70 & 31.14 & 66.36 & 28.06 & 88.51 & 467 \\
         \cline{1-11}
         INT8+4 & 98.97 & 24.18 & 97.69 & 15.12 & 95.60 & 29.31 & 69.48 & 31.39 & 88.53 & 489 \\
         \cline{1-11}
         INT4+8 & 98.99 & 22.71 & 97.56 & 12.31 & 96.90 & 28.41 & 70.39 & 36.57 & 87.76 & 215  \\
         \cline{1-11}
         INT4+4 & 97.40 & 31.49 & 94.97 & 12.33 & 93.80 & 23.72 & 65.87 & 32.46 & 86.01 & 186  \\
        \hline 

    \end{tabular}
    \label{tab:training}
\end{table*}

\Cref{tab:training} summarizes the results of our study. First, we observe consistent trends in the distribution of per-exit accuracy and exit ratios across all experiments. The networks exhibit high confidence in relatively easy samples, leading to high per-exit accuracy at intermediate exit points and a 63\% reduction in computation for nearly 40\% of cases (corresponding to samples exiting at points 1 and 2). However, more feature extraction is required in deeper layers for harder samples. Additionally, the accuracy at the last exit tends to be lower due to the increased difficulty of these samples. Hence, we verify the efficiency of EENN across all quantization levels.
Second, across various quantization configurations, we observe notable differences in per-exit accuracy and exit-ratio distributions. Although the full-precision and 8-bit precision models exhibit only a small difference in average accuracy, the full-precision model benefits from increased expressivity, with the first exit achieving an exit ratio of 34.21\%, compared to just under 30\% for the quantized models. This results in more frequent late exits in the quantized models. When both the backbone and early exits are quantized to 4 bits, significant hardware cost reductions are achieved, but the average accuracy suffers due to lower precision. In contrast, mixed quantization schemes provide a balanced trade-off between soft (e.g., 8-bit) and hard (e.g., 4-bit) homogeneous quantization. This approach minimizes the impact on average accuracy while still achieving notable hardware cost savings. Specifically, quantizing the backbone to 4 bits reduces the computational burden of heavy operations (e.g., backbone convolutions) while allowing the exits to compensate for accuracy loss due to their larger capacity for information encoding.

\begin{figure}[t]
    \centering
    \includegraphics[width=1\linewidth]{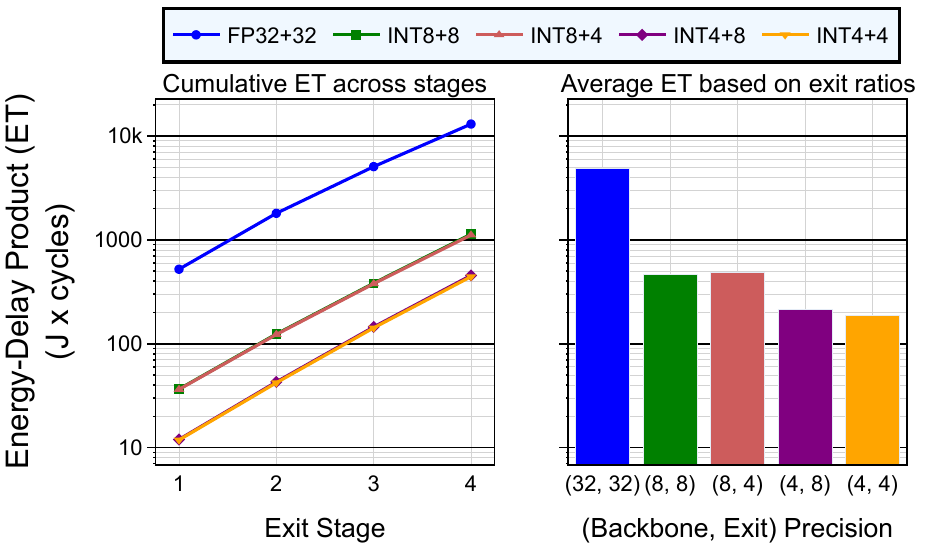}
    \caption{The cumulative (left) and average (right) energy-delay product ET at different stages for differently quantized models. The ET\_{avg} is drastically reduced thanks to the early exiting at early stages.}
    \label{fig:precision-et}

\end{figure}

Figure~\ref{fig:precision-et} details the breakdown of the energy-delay product (ET) in different exit stages (left) and shows $ET_{\text{avg}}$ for different precisions (right). Note the logarithmic scale of the Y-axis, which demonstrates that $ET_{\text{avg}}$ is significantly lower than the worst case due to the high exit ratios at earlier stages. Moreover, the reduction in $ET_{\text{avg}}$ for reduced precision is shown. Interestingly, we note that the model quantized with 8 bits for the backbone and 4 bits for the exits exhibits a slightly worse $ET_{\text{avg}}$ than the 8-bit + 8-bit quantized model, despite being more aggressively quantized. This is due to lower exit ratios at the early stages. Hence, quantization and early exit can interact in complex ways, potentially leading to significant performance differences, as observed in mixed-quantization models.

\subsection{Impact of Mounting Points}

\begin{figure}[t]
    \centering
    \includegraphics[width=1\linewidth]{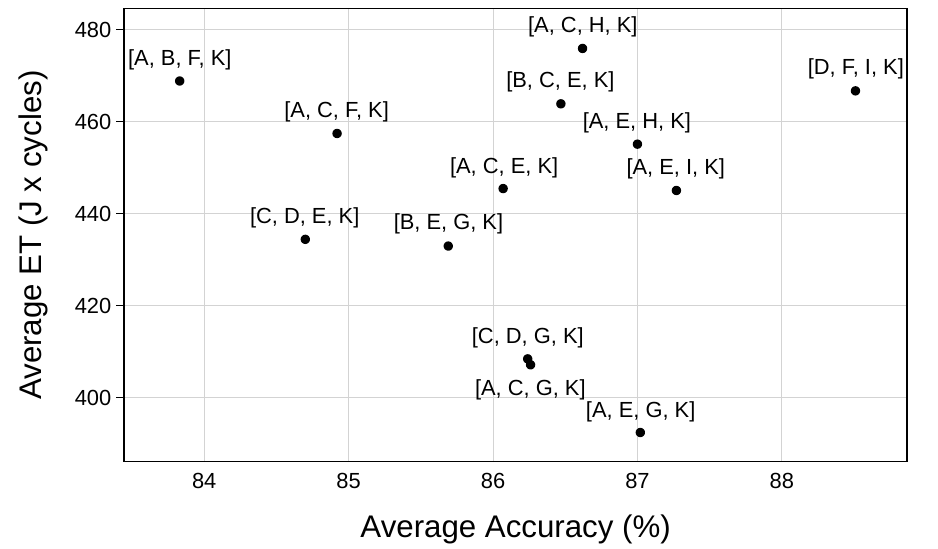}
    \caption{Family of 4-exit models in INT8+8 configuration with identical backbone and exits topologies. Each model has different exit points, identified by the exit indices (cf., Table 1 in Appendix A).}
    \label{fig:4_exits}
\end{figure}

Next, we study the interplay between early exit mounting points and hardware performance. For this, we use 8-bit quantization for both the backbone and the exits. \Cref{fig:4_exits} shows the average accuracy and ET for models with different mounting points of the four exits (denoted by four letters; cf. Table 1 in Appendix A). We observe that some architectures become severely unfit (e.g., model [A, B, F, K], which achieves low accuracy and high hardware cost), while others are more promising, offering a reasonable balance between accuracy and ET, yielding a subset of efficient networks to choose from (e.g., model [A, E, G, K]). However, no obvious patterns emerge for efficiently designing EENN. For instance, if we examine the four models [A, C, x, K], where x can be E, F, G, or H, we find significant differences in ET and accuracy between the models after minimal changes in the third exit point placement. Taking x to be E results in high accuracy and low ET, while moving it by a single block to F severely degrades performance.

\Cref{fig:distrib_02} presents the execution time (ET) at different exit points for a subset of the best-performing models depicted in \Cref{fig:4_exits}. The results indicate that most of the degradation in ET comes from the difference in the placement of the third exit, while differences in the other exits are minor. This showcases the complex nature of the interactions between the placement of exit points and the hardware, which may have a significant impact on energy efficiency in some cases while being relatively harmless in others.
Another example is the behavior of [A, C, G, K] and [C, D, G, K], which achieve similar performance despite having different first and second exit points. These differences could be explained by complex interactions between the layers' shapes and the dataflows of the hardware cores, as different mounting points can have different activation and channel dimensions. For instance, if the dimensions are not a power of 2, this may lead to underutilization of the compute array.

\begin{figure}[t]
    \centering
    \includegraphics[width=1\linewidth]{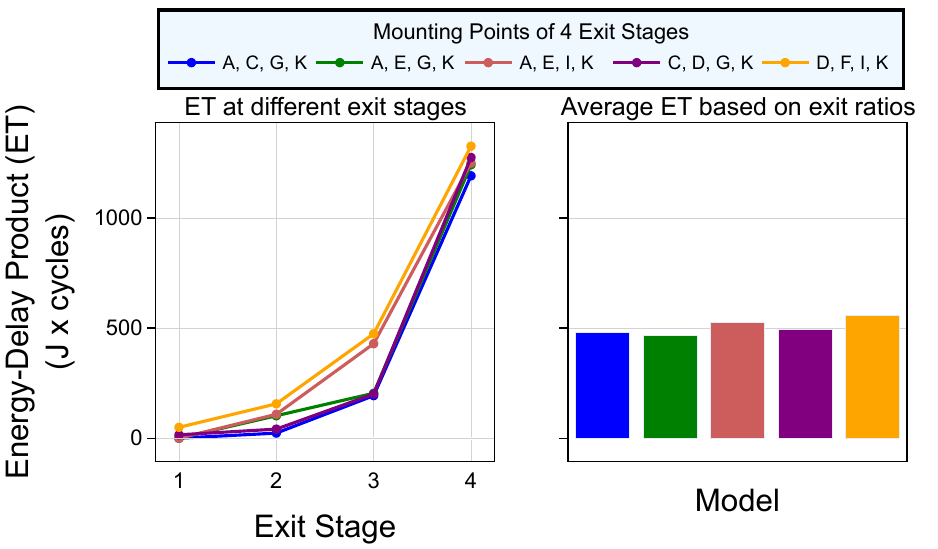}
    \caption{The cumulative (left) and average (right) energy-delay product ET at different stages for different mounting points with identical backbone and exit architecture. The ET varies due to exit ratio differences and tensor dimension mismatches with the accelerator dataflows.}
    \label{fig:distrib_02}

\end{figure}

\subsection{Impact of Accelerator Types}

\begin{figure}[t]
    \centering
    \includegraphics[width=1\linewidth]{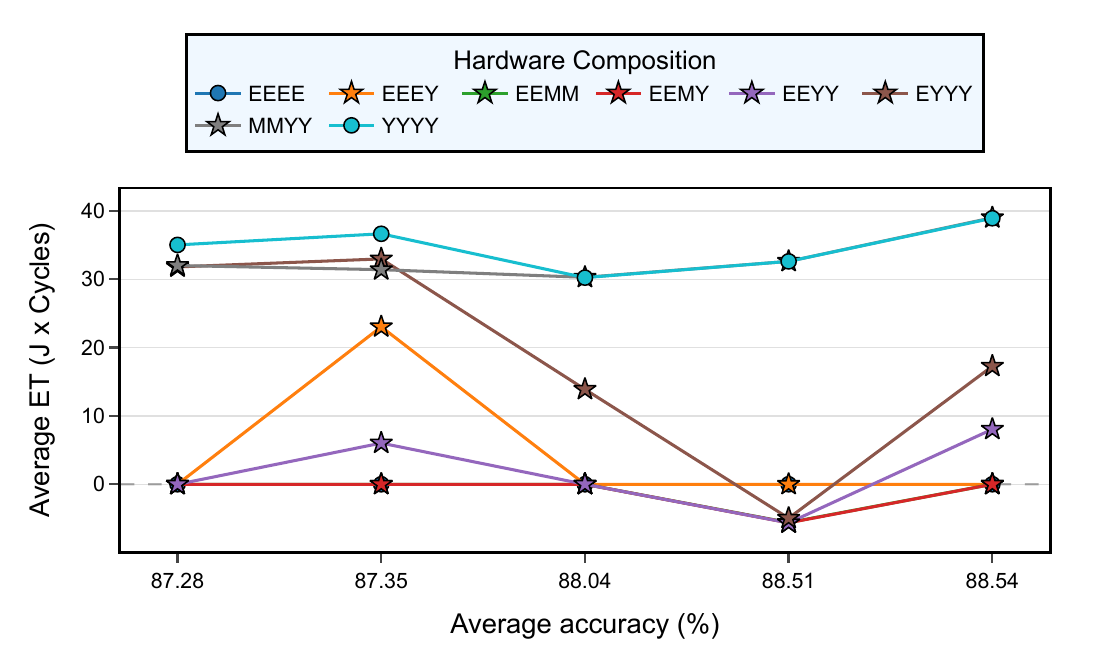}
    \caption{Difference in $\mathrm{ET}_{\mathrm{avg}}$ relative to the homogeneous EEEE configuration versus $\mathrm{ACC}_{\mathrm{avg}}$ for five representative EENN architectures evaluated on eight quad-core accelerator configurations of identical compute and memory budgets, composed of Edge TPU-like (E), Meta-prototype-like (M), and Eyeriss-like (Y) cores. The MMMM configuration is omitted for clarity due to its substantially higher $\mathrm{ET}_{\mathrm{avg}}$.}

    \label{fig:case_study}    
\end{figure}

To illustrate the impact of hardware composition on EENN deployment, we evaluate the deployment of five random EENN architectures, characterized by different prediction accuracies (X-axis in \Cref{fig:case_study}) on eight quad-core accelerator configurations of identical compute and memory budgets, composed of three processing elements: an Edge TPU-like core (E), a Meta-prototype-like core (M), and an Eyeriss-like core (Y) (cf. Appendix A) under an 8-bit quantization scheme. To facilitate comparison across hardware platforms, the homogeneous EEEE configuration is used as the reference baseline. Consequently, the Y-axis reports the difference in average execution time ($\Delta\mathrm{ET}_{\mathrm{avg}}$) with respect to the EEEE platform, where a value of zero corresponds to the baseline and positive values indicate higher energy-delay product.

Two major observations can be drawn from this experiment. First, no single hardware configuration consistently provides the best performance across all evaluated architectures. While the homogeneous EEEE platform achieves the lowest ET for several models, heterogeneous configurations become preferable for others. For example, for the architecture achieving 88.51\% accuracy, the heterogeneous EEMM platform outperforms the homogeneous EEEE configuration. This demonstrates that even modest changes in the neural architecture can alter the most suitable hardware platform. Second, the relative ranking of accelerator configurations changes from one architecture to another, revealing complex and non-linear interactions between network architecture and hardware characteristics. These interactions arise from differences in computation patterns, memory accesses, and workload distribution, making the deployment performance difficult to predict through manual design decisions alone. As a result, these experiments confirm that mapping-aware evaluation is necessary in searching optimal deployable EENN rather than optional.

We therefore propose a deployment-aware co-design framework that jointly optimizes these design dimensions through NAS. By systematically exploring the coupled algorithm-hardware design space, our approach identifies efficient EENN deployments without relying on exhaustive manual tuning, enabling scalable optimization across heterogeneous edge platforms.

\section{Deployment-aware Co-Design Framework for EENN}
\label{approach}

In this section, we present our deployment-aware co-design framework for quantized EENNs targeting edge hardware architectures. We begin by formulating the constrained multi-objective optimization problem that captures the interplay between architecture, quantization, and hardware mapping. We then describe the methodological steps through which the framework systematically explores this coupled design space to identify efficient EENN configurations.

\subsection{Problem Formulation}
\begin{figure*}[t]
    \centering
    \includegraphics[width=0.75\textwidth]{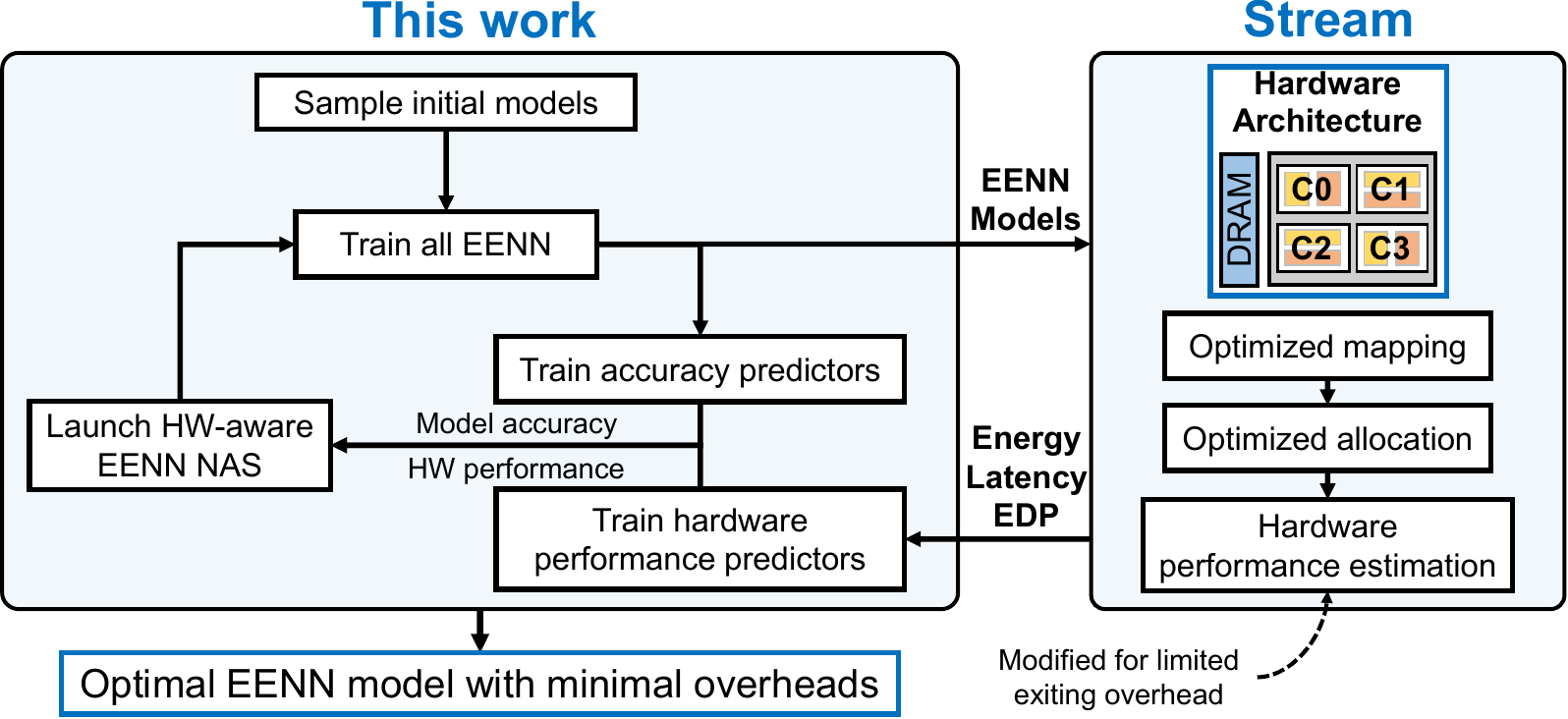}
    \caption{Overview of the deployment-aware co-design framework, integrating quantization-aware training and analytical hardware cost estimation via the Stream~\cite{stream} design space exploration platform to evaluate energy and latency at different exit configurations.}

    \label{fig:overview}
\end{figure*}

Our objective is to transform a DL model into an efficiently deployable EENN tailored to multi-core edge accelerators. The backbone may serve diverse applications, including image classification~\cite{shallow_deep_nets}, audio denoising~\cite{mlsp}, or transformer-based sentiment analysis~\cite{bert_ee}. Rather than treating early exits as architectural add-ons, we formulate their integration as a deployment-aware co-design problem that jointly accounts for predictive performance and hardware efficiency.

Specifically, we seek to identify an EENN configuration that balances accuracy and energy-latency cost under realistic hardware constraints. In addition to optimizing these conflicting objectives, we impose structural constraints that reflect practical deployment considerations. First, the computational overhead introduced by each intermediate exit must remain bounded relative to the remaining backbone computation. Second, the network must effectively exploit early exiting by limiting the proportion of samples that propagate to the final classifier.

Formally, the EENN design problem is expressed as the following constrained multi-objective optimization problem:
\begin{equation}
\begin{aligned}
\label{nas_prob}
       &\max_{s \in \mathcal{S}} \ \text{ACC\_{avg}}(s)\\
       &\min_{s \in \mathcal{S}} \ \text{ET\_{avg}}(s)  \\
    \text{s.t.} \ &\begin{cases}
            \text{OH}_i(s) \leq \theta, \ \forall i \in \{1, \dots, m\} \\
            \text{ER}_m(s) \leq \mu
        \end{cases}
\end{aligned}
\end{equation}
where $s$ denotes an architecture from the search space $\mathcal{S}$, and $m$ is the total number of exits, including the final classifier. The term $\text{OH}_i(s)$ is the ratio between the additional ET cost incurred by exit point $i$ and the ET of the backbone layers up to the next exit point (e.g., in \Cref{fig:eenn}, $\text{OH}_1(s)$ corresponds to the ratio of the ET of Exit 1 to the ET of Backbone 2); $\theta$ is the user-defined threshold on the overhead, and $\text{ER}_m$ is the exit ratio of the last exit, which is upper-bounded by a user-defined threshold $\mu$. Hence, our main goal is to optimize two conflicting objectives, namely EENN's average accuracy and energy-latency product, within the constrained search space of networks with bounded intermediate exit overhead and effective early exiting.

The resulting formulation captures the inherent trade-off between predictive accuracy and hardware efficiency while explicitly constraining architectural overhead and dynamic inference behavior. This perspective treats EENN deployment as a joint hardware-algorithm optimization problem rather than a purely architectural search task.

\subsection{Search Space Design}

The design space of deployment-aware EENN configurations is defined by the structural properties that directly influence both predictive behavior and hardware efficiency. These include the number of intermediate exits, their placement along the backbone, the architectural configuration of each exit (e.g., depth and width of the classifier block), and the quantization level applied to backbone and exit components. Consistent with our objective of augmenting existing models rather than redesigning them, the backbone architecture is treated as fixed and is therefore excluded from the search space. This reflects practical deployment scenarios where the backbone is predetermined by the target application or an existing pretrained model, while the design flexibility lies in the early-exit mechanism.

Given a fixed DL backbone, let $H$ denote the maximum number of admissible intermediate exits (excluding the final classifier), $q$ the set of possible quantization levels, and $p$ the set of candidate exit architectures. For a configuration with $k$ intermediate exits, the number of architectural and quantization combinations is $p^{k+1}$ and $q^{k+1}$, respectively, accounting for the final classifier as well. Consequently, the total number of candidate configurations is $\sum_{k=0}^{H} \binom{H}{k} (pq)^{k+1},$ which, by the binomial theorem, simplifies to $pq(1+pq)^H$.
This exponential growth illustrates the combinatorial nature of the joint architectural-quantization design space. Since exhaustive enumeration is computationally prohibitive, guided exploration strategies are required. In our framework, we adopt a genetic algorithm (GA) due to its flexibility in handling discrete, structured variables and its suitability for constrained multi-objective optimization \cite{survey_hw_nas}. The GA enables efficient navigation of the coupled design space while respecting deployment-driven constraints introduced in the previous subsection.

\subsection{Automatic Search Procedure} 

The proposed deployment-aware co-design framework, illustrated in \Cref{fig:overview}, proceeds through four main stages.\\
({\em Step 1}) The process begins by sampling various EENN architectures to form an initial set of candidate models, denoted $\mathcal{S}^0$, that comply with the overhead constraint $\theta$.\\
({\em Step 2}) Each architecture in $\mathcal{S}^0$ is trained using quantization-aware training and evaluated on the Stream platform to obtain its corresponding accuracy and energy-delay product metrics. For each trained model, we apply the last exit ratio constraint $\mu$ and remove those that do not satisfy it.\\
({\em Step 3}) The metrics collected from each model constitute a labeled dataset: $\mathcal{P}^0 =  \{\left(s, \text{SM}(s),  \text{HM}(s)\right) \ | \ s \in \mathcal{S}^0\} $, where each architecture is mapped to its respective software metrics defined as $\text{SM}(s)=(\text{ACC\_{avg}}(s), \text{ER}_m(s))$, and hardware metrics defined as $\text{HM}(s)=(\text{ET\_avg}(s), \text{OH}_i(s))$. The dataset $\mathcal{P}^0$ is then used to train surrogate predictors that rapidly estimate the software and hardware performance metrics of candidate architectures during NAS, significantly reducing the computational cost of the search. \\
({\em Step 4}) Finally, the GA triggers an automatic search over multiple generations to generate better architectures. One generation proceeds as follows:
\begin{itemize}
    \item Based on the previously trained predictors, the accuracy and ET are predicted for the parent architectures in the initial population.
    \item The population is ranked according to accuracy, and the top $2N$ parents are shortlisted. These are then ranked by ET value to retain the best $N$ architectures for applying the GA operators. 
    \item The GA applies mutation and crossover operators to the chromosomes of the parent architectures, which describe the EENN configuration (i.e., mounting points, depth of each early-exit network, quantization level, etc.).
    \item Each generation of GA offspring is filtered to remove models that do not satisfy the overhead constraint.
    \item The new generation forms a new population, to which the same steps are applied.
\end{itemize}
At the end of the GA process, a new set $\mathcal{S}^1$ is created, consisting of the best $N$ offspring and their ancestors from $\mathcal{S}^0$. Steps 2 to 4 are then repeated on $\mathcal{S}^1$, and the process continues for a finite number of iterations or until a desired balance between accuracy and ET cost is reached.
However, previously trained architectures are not retrained during Step 2, and the training of predictors in Step 3 is done using the combined dataset: $\mathcal{P}^k = \bigcup_{i=0}^k \mathcal{P}^{i}$. 

In conventional NAS, performance predictors are trained only once from a large set $\mathcal{P}^0$ (e.g., four orders of magnitude in \cite{enas4d}), yielding strong predictors. This approach results in a sequential NAS pipeline where $\mathcal{S}^1$ is not trained, and the best-performing architecture is retained, thereby ending the search.
However, our co-design framework aims to combine optimality (i.e., finding the best architecture) and efficiency (i.e., a shorter search time compared to conventional NAS). Hence, we adopt a more efficient approach based on progressive weak predictors, originally proposed in \cite{weaknas}. The idea is to alternate between training predictors using increasingly large datasets (i.e., $\mathcal{P}^k$) and running the GA process.
Therefore, our co-design approach can be formulated as the generation of the two sets $\mathcal{P}^{k}$ and $\mathcal{S}^{k+1}$ at iteration $k$ as follows:
\begin{equation}
\label{fit_pred}
\begin{cases}
\mathcal{P}^{k} = \left\{ (s, \text{SM}(s),  \text{HM}(s) ) \ \vert \ s \in S^{k} \right\} \\
S^{k+1} = \text{Top\_GA}_N(\mathcal{S}^{k}) \cup \mathcal{S}^{k}
\end{cases}
\end{equation}

\noindent where $\text{Top\_GA}_N$ a function that returns the best $N$ GA offspring architectures from an input set of parent architectures $\mathcal{S}^{k}$ ordered based on the estimated accuracy and ET cost predictors.

\section{Experimental Evaluation}
\label{exprms}
In this section, we experimentally evaluate the proposed deployment-aware co-design framework on an image classification task representative of edge deployment scenarios.

\subsection{Experimental Setup}
Similarly to \Cref{study}, we use an image classification task on CIFAR-10 to evaluate our co-design framework on a quad-core edge TPU for deploying an 8-bit quantized EENN (cf. Appendix A). Our search space is built on a fixed 12-block-deep MobileNetV2 backbone. Early exits can be mounted in the positions denoted by letters in Table 1 in Appendix A. The search parameters are the depth, position, and number of early exits, which are encoded using a one-hot representation.
The exit-point classifier consists of max-pooling operators that reduce the input tensors' height and width to 4×4 tensors with the same number of channels, as done in~\cite{shallow_deep_nets}, followed by one or two linear layers with ReLU6 activation functions. We constrain newly generated EENN architectures according to $\theta = 50\%$ and $\mu = 50\%$.

\subsection{Proposed Framework Results}

\begin{figure}[t]
    \centering
    \includegraphics[width=1\linewidth]{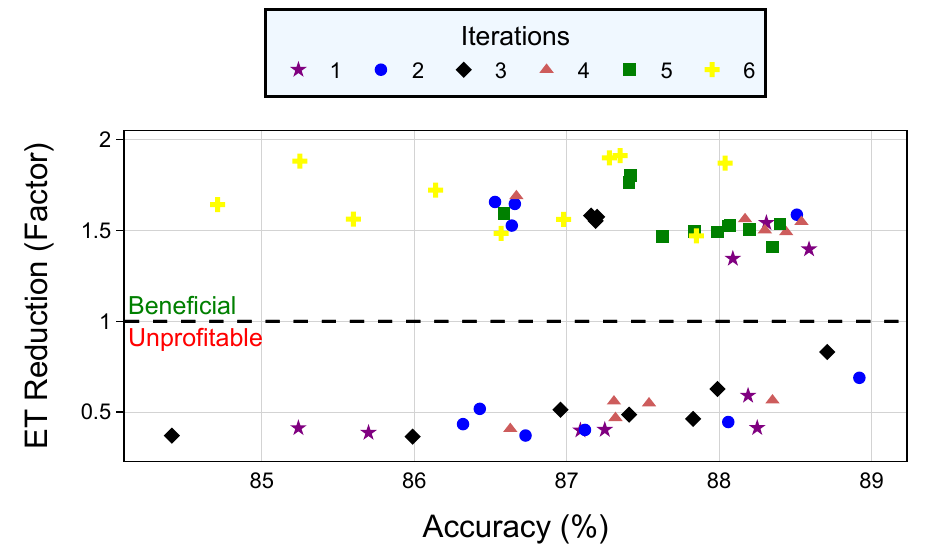}
    \caption{The development of the search with respect to accuracy and EENN ET reduction (i.e. ratio of ET\_{avg} and the respective static architecture ET, without early exits). More exploration tends to yield architectures with high accuracy and high ET reduction.}
    \label{fig:dynamic_hw}

\end{figure}

\Cref{fig:dynamic_hw} presents the reduction in the ET\_{avg} relative to the static backbone (where all samples exit at the final stage without considering intermediate exit overhead) on the y-axis, and the x-axis shows ACC\_{avg}. The figure reveals that our framework explores the search space covering inefficient models with low accuracy and high ET\_avg and tends to exploit local regions where both objectives are well-balanced as iterations increase. With more iterations, the inherent random nature of the search is reduced, thereby avoiding low-outcome regions of the search space and exploiting others that are more promising for better trade-offs (e.g., iteration 5). However, due to the stochasticity of the GA operators, a few inefficient architectures can still be encountered (e.g., iteration 6). Nevertheless, efficient architectures can be found with as few as 5 NAS iterations, which can be explained by the presence of constraints on the overhead and exit ratios that were added to our search space. 

Furthermore, we can also separate the models based on their profitability, i.e., when early exiting brings a gain compared to the static backbone (see dotted line in \Cref{fig:dynamic_hw}). Across the entire dataset, the NAS process identifies architectures that are notably more efficient than their static counterparts. These optimized architectures can achieve improvements exceeding $50\%$. This highlights the potential benefits of enhancing conventional networks with early exiting mechanisms, particularly when tailored to specific use-case objectives. 

\begin{figure}[t]
    \centering
    \includegraphics[width=1\linewidth]{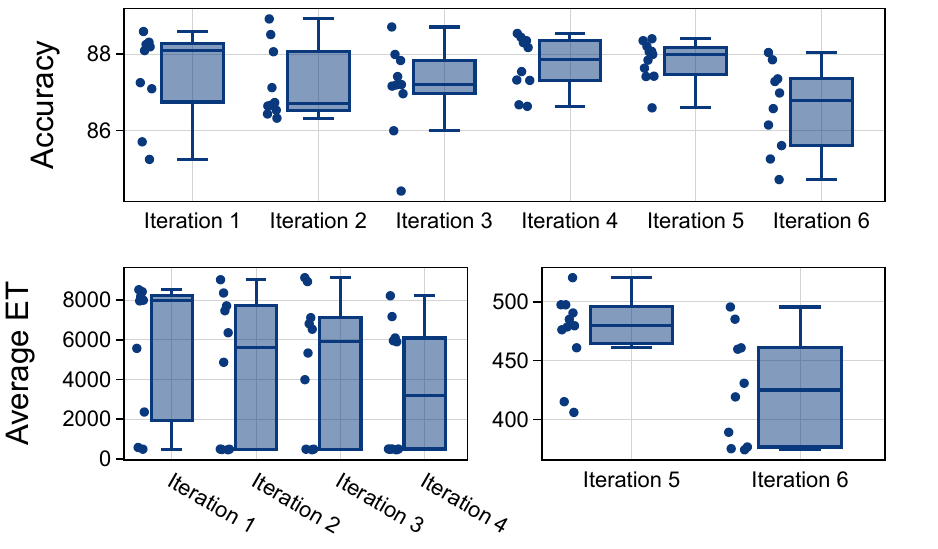}
    \caption{Distribution over EENN models in terms of Accuracy (in \%) and Average ET (in J x cycles) across NAS iterations.}
    \label{fig:whisker}
\end{figure}

To gain deeper insights into the evolution of models, we visualize the distributions of accuracy and ET\_avg across NAS iterations in \Cref{fig:whisker}. We observe in the first iterations a large exploration of the search space with different architectures of variable performance (e.g., models with an ET\_avg exceeding 8000 J x cycles). As the number of iterations grows, the NAS capitalizes on previously selected best-performing architectures thus increasing the overall mean accuracy and decreasing the overall mean ET\_avg of the found models. Moreover, the balancing effect of our conflictual objectives can be observed from iterations 5 and 6, where the mean accuracy was slightly reduced to gain in terms of ET. This is mainly due to an implicit higher weight given to ET\_avg from the ranking of generated architectures after fairly stabilizing accuracy.

\subsection{Analysis of Efficient Architectures}

\begin{figure}[t]
    \centering
    \includegraphics[width=1\linewidth]{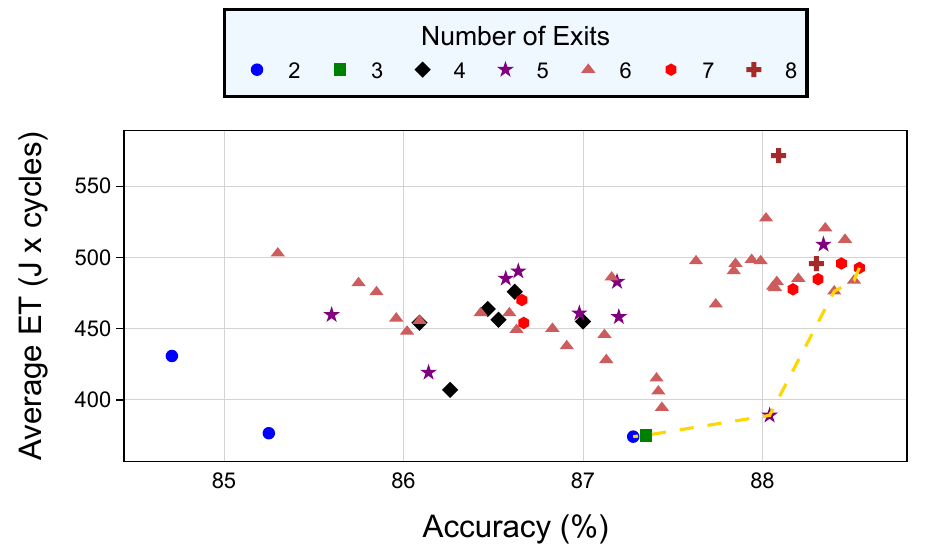}
    \caption{Subset of efficient architectures (where $\text{ET\_{avg}} < 1000$) from the NAS showing the Pareto front (gold line) with varying number of exits.}
    \label{fig:nas_accuracy_avg_et}    
\end{figure}

In this section, we analyze a subset of efficient architectures explored during the NAS process with $\text{ET}_{\text{avg}} < 1000$.
As seen in \Cref{fig:nas_accuracy_avg_et}, there is a notable concentration of architectures with a high number of exits. This observation can be explained by the fact that a greater number of exits generally improves performance, as it provides more opportunities for input samples to terminate earlier in the network, thereby reducing computational costs. Specifically, additional exits allow for the early termination of input samples that can be processed more quickly, leading to lower overall resource usage.
However, the benefits of increasing exits are not without trade-offs. A greater number of exits may negatively impact model accuracy due to the amplification of the cascading effect in gradient propagation during backpropagation. As the number of exits increases, the gradients may become more dispersed, potentially reducing the effectiveness of weight updates. Additionally, a higher exit count can impose greater demands on the hardware, particularly by increasing reliance on max-pooling operations and inter-core data transfers, which can reduce computational efficiency.
Furthermore, we observe from the landscape that accuracy improves with a higher number of exits, albeit at the expense of a higher $\text{ET}_{\text{avg}}$. Indeed, the Pareto-optimal architectures (gold line in \Cref{fig:nas_accuracy_avg_et}) follow the same trend and prove that our NAS strives to balance the trade-off between accuracy and computational efficiency.

\begin{table}[b]
    \centering
    \caption{Comparison of efficient models in terms of accuracy and MAC operations from previous EENN-specific NAS studies.}
    \begin{tabular}{c|c|c|c}
    \hline
    \textbf{Method} & \textbf{Precision} & \textbf{Accuracy (\%)} & \textbf{MAC Reduction (\%)} \\
    \hline 
    EDANAS \cite{edanas} & FP32 & 81.10 & 36.79 \\
    \hline
    NACHOS \cite{nachos}  & FP32 & 72.65 & 58.99 \\
    \hline 
    \textbf{Ours} & INT8 & \textbf{88.04}  & \textbf{56.46} \\
    \hline
    \end{tabular}
    
    \label{tab:compare}
\end{table}

Even though our framework brings new constraints to the training (i.e., quantization, last exit ratio) and optimizes more complex hardware-related metrics (i.e., energy and latency via DSE), we intend to compare our efficient models with similar studies. We compare against EDANAS \cite{edanas} and NACHOS \cite{nachos}, with which we have close initial conditions, where EENN backbones are built on MobileNets and trained on CIFAR-10. It is worth noting three major differences: (1) their hardware cost is modeled with the number of MAC operations, (2) their training procedure is based on \cite{differential_branching} whereas our models are trained using linear scalarization under quantization, and (3) their backbone parameters (e.g., kernel size, depth) are extra dimensions of the search space, contrarily to our case where the backbone is fixed. In \Cref{tab:compare}, we report the accuracy and MAC reduction of the best models from EDANAS and NACHOS, along with a Pareto optimal model from our NAS (c.f., star marker in the Pareto front gold line in \Cref{fig:nas_accuracy_avg_et}). We show that our framework is competitive, yielding highly accurate and MAC-efficient architectures. The added constraints in our NAS helped the search be more effective in finding architectures that are both hardware-friendly (i.e., overhead) and that benefit well from early exiting (i.e., last exit ratio). Overall, thanks to the constraints we impose on our NAS framework, we are able to find well-adapted EENN models with reduced design time for fast deployment. Furthermore, with a rich Pareto front, the framework facilitates the identification of the most suitable model aligned with real-world performance and resource requirements.

\section{Conclusion}
\label{conclusion}
 
This work presented a deployment-aware hardware-algorithm co-design framework for EENNs targeting heterogeneous multi-core edge accelerators. By jointly optimizing exit configuration, quantization-aware training, and hardware mapping within a unified NAS process, the proposed framework identifies architectures that achieve a superior trade-off between predictive accuracy and hardware efficiency. Experimental results demonstrate that considering deployment decisions during the search substantially improves the energy-latency product compared with conventional architecture-centric optimization. More broadly, our results highlight that the efficiency of dynamic neural networks depends not only on their architecture but also on how they are deployed on the target hardware. Integrating hardware mapping into the optimization process therefore provides a more faithful assessment of candidate architectures and enables solutions better suited for modern edge accelerators. Future work will investigate finer-grained dynamic inference mechanisms, alternative early-exit decision policies, and the extension of the proposed framework to a broader range of neural network architectures and heterogeneous accelerator platforms.

\section*{Appendix}
In this appendix, we present further details about the backbone architecture, the accelerator architectures, and the adopted quantization-aware training strategy.

\subsection{Model \& Hardware Base Architectures}
\label{appendix_implementation}

\noindent \textbf{EENN Backbone Model:} We adopt MobileNetV2 as the backbone network due to its widespread use in previous EENN NAS studies \cite{edanas,nachos}. Specifically, we employ the modified MobileNetV2 architecture shown in \Cref{tab:mbv2}. All depthwise convolutions use a kernel size of $3$, padding of $1$, and an expansion factor of $6$. Each bottleneck consists of the number of repeated blocks, where each block is composed of three convolutional layers. The number of channels is expanded during the second convolution of each block. Early exits may be inserted after every block within bottlenecks A-J, providing ten candidate exit locations, while the final classifier is always placed at the end (position K).

\begin{table}[h!]
    \centering
    \caption{MobileNetV2 with Early Exits Mounting Points}
    \begin{tabular}{c|c|c|c|c}
    \hline
        \textbf{Operator} & \textbf{Repetition} & \textbf{Exit Index} & \textbf{Channels} & \textbf{Strides} \\
        \hline \hline
        conv2d & 1 & - & 32 & 1 \\
        \hline
        bottleneck & 1 & - & 16 & 1 \\
        \hline
        bottleneck & 2 & A, B & 24 & 1  \\
        \hline
        bottleneck & 2 & C, D & 32 & 1  \\
        \hline
        bottleneck & 2 & E, F  & 64 & 2  \\
        \hline
        bottleneck & 2 & G, H & 96 & 1  \\
        \hline
        bottleneck & 2 & I, J & 160 & 2  \\
        \hline
        bottleneck & 1 & K & 320 & 1 \\
        \hline
    \end{tabular}
    \label{tab:mbv2}
\end{table}

\noindent \textbf{Hardware Accelerators:} We consider three representative accelerator architectures modeled in the Stream framework: a TPU-like \cite{coral_edge}, an Eyeriss-like \cite{eyeriss}, and a Meta-prototype \cite{meta_proto} accelerator. While the three architectures share the same overall hardware budget, they differ in the organization of their register files and their connectivity to the compute array. In the TPU-like and Meta-prototype accelerators, a single register file read is broadcast to sixteen multipliers, whereas in the Eyeriss-like accelerator each multiplier is equipped with its own private register file. Aside from this architectural difference, all three accelerators feature approximately 1024 MAC units, 2 MiB of on-chip memory, and identical MAC energy characteristics. Additional dedicated cores handle pooling operations and SIMD instructions required for element-wise additions and multiplications. The compute cores communicate with off-chip memory through a 64-bit/cycle interface, while memory energy costs are scaled from \cite{horowitz_scaling}. All accelerator models support flexible numerical precision for both computation and memory accesses, enabling the evaluation of quantized EENN deployments. Unless otherwise specified, all experiments are conducted using a quad-core edge-scale TPU-like accelerator.

\subsection{Quantization-aware Training of EENN}
\label{appendix_train}


Each EENN is trained by jointly optimizing the losses associated with its $m$ intermediate classifiers. The exit-specific losses are aggregated into a single objective using weighted linear scalarization,

\begin{equation}
\mathcal{L}_{LS}=\sum_{i=1}^{m}\lambda_i\mathcal{L}_i,
\end{equation}

\noindent where $\mathcal{L}_i$ denotes the loss associated with the $i^{th}$ exit and $\lambda_i$ its corresponding weight. Following previous EENN works \cite{shallow_deep_nets,s2dnas}, all loss weights are set to one. All models are trained for $100$ epochs using mini-batch gradient descent with a learning rate of $10^{-3}$, momentum of $0.9$, weight decay of $5\times10^{-4}$, and a batch size of $128$.

To account for the low-precision arithmetic used by modern edge accelerators, all candidate EENNs are trained using QAT. During training, both weights and activations are linearly quantized, allowing the network to adapt to quantization effects before deployment. Specifically, each real-valued parameter $r$ is clipped to the range $[-c,c]$ and quantized into $b$ bits according to:
\begin{equation}
Q(r)=\text{round} \left (
\frac{\max\left(-c,\min(r,c)\right)}{s}
\right) \times s,
\end{equation}
where
\begin{equation}
s=\frac{c}{2^{b-1}-1}.
\end{equation}

Following \cite{ofa_quantized}, the clipping threshold $c$ is selected independently for each layer by minimizing the KL-divergence between the original and quantized value distributions.

\bibliographystyle{IEEEtran}
\bibliography{refs}

\end{document}